%% file: paper_arXiv_v1.tex
\documentclass[%
 reprint,
 superscriptaddress,
 amsmath,
 amssymb,
 aps,
 prl,
]{revtex4-2}

\usepackage{graphicx}
\usepackage{dcolumn}
\usepackage{bm}
\usepackage{hyperref}
\hypersetup{
    colorlinks=true,
    linkcolor=blue!50!black,
    citecolor=blue!50!black,
    urlcolor=blue!50!black
}
\usepackage[T5,T1]{fontenc}

\usepackage{caption}
\usepackage{subcaption}
\usepackage{tabularray}
\UseTblrLibrary{booktabs}
\UseTblrLibrary{siunitx}
\usepackage{xcolor}
\usepackage{siunitx}
\usepackage{csquotes}
\usepackage{xfrac}
\usepackage{placeins}
\usepackage[nameinlink]{cleveref}
\usepackage{orcidlink}
\usepackage{titlesec}
\usepackage{float}
\usepackage{mathtools}

\DeclarePairedDelimiter{\abs}{\lvert}{\rvert}

\crefname{figure}{Fig.}{Figs.}
\crefname{equation}{Eq.}{Eqs.}

\newif\ifarxiv
\arxivtrue        

\begin{document}

\preprint{}

\newcommand{\papertitle}{Measurement of the Neutrino Spectrum along the Galactic Plane with IceCube}
\title{\papertitle} 

\input{authorlist.tex}

\date{\today}

\collaboration{IceCube Collaboration}
\noaffiliation

\begin{abstract}
We present the first measurement of the variation of the high-energy neutrino flux along the Galactic plane
using 10 years of neutrino-induced particle-shower data from the IceCube Neutrino Observatory.
In this Letter, we fit both the neutrino flux normalization and the spectral index, in multiple regions along the plane, independent of model expectations. 
The neutrino flux per solid angle in the inner Galactic region ($|l|<\qty{30}{\degree}, |b| < \qty{8}{\degree}$)
is approximately five times higher than in the outer Galactic plane ($|l|>\qty{60}{\degree}, |b| < \qty{8}{\degree}$) at $E_\nu=\qty{5}{\tera\electronvolt}$.
The spectral indices in the inner and outer region are
$\gamma=\num{2.76(18:15)}$
and 
$\gamma = \num{3.13(57:37)} $, respectively.
Additionally, we compare the same region of Galactic emission measured in $\unit{\tera\electronvolt}$ gamma rays 
by LHAASO ($\qty{15}{\degree}<l<\qty{125}{\degree}$) to neutrinos, 
and conclude that the measurements are compatible within uncertainties, under the assumption of hadronic gamma-ray production.
\end{abstract}

{
\hypersetup{pdfborder={0 0 0}}
\maketitle
}

\titleformat{\section}[runin]
  {\normalfont\bfseries}
  {}
  {0pt}
  {}
  [\,---]
\titlespacing{\section}{0pt}{0pt}{.16667em} 

\begin{figure*}
\centering
\begin{subfigure}{.49\textwidth}
\includegraphics[width=1\textwidth]{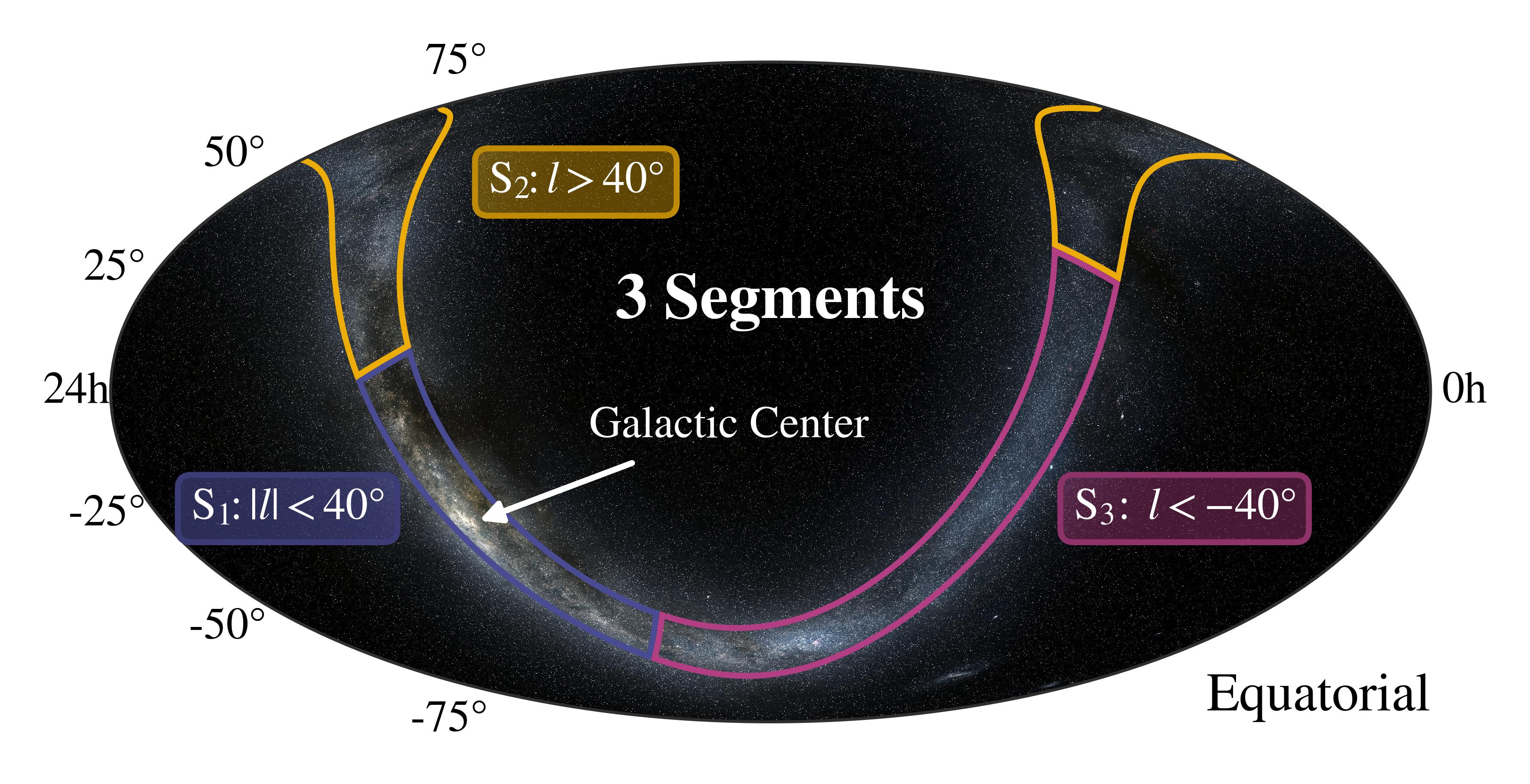}
\caption{One example of the eight tested segmentation schemes of the GP in equatorial coordinates. This generic scheme divides the GP into an \enquote{inner}, \enquote{right}, and \enquote{left} region.}\label{fig:generic-3}
\end{subfigure}
\begin{subfigure}{.49\textwidth}
\includegraphics[width=1\textwidth]{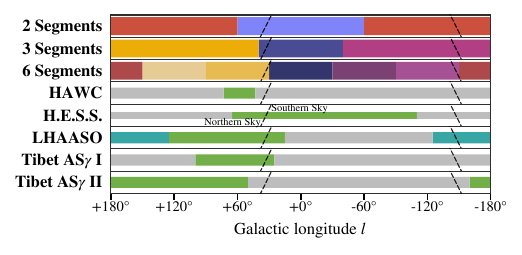}
\caption{Each box depicts \qty[print-implicit-plus]{+180}{\degree} to \qty{-180}{\degree} in Galactic longitude and \qty{+-10}{\degree} in Galactic latitude. The size and locations of the tested regions are depicted by the colored bars.}\label{fig:overview}
\end{subfigure}
\caption{Illustration of the analysis method. (\subref{fig:generic-3}) shows one segmentation scheme in equatorial coordinates on top of an optical image of the Milky Way~\cite{Mellinger2009} and (\subref{fig:overview}) shows all tested Galactic regions in Galactic coordinates.}\label{fig:segmentation}
\end{figure*}

\section{Introduction}\label{sec1}
For many areas of astronomy, the Milky Way is the brightest region in
the sky.
It has been observed throughout the electromagnetic spectrum from radio waves to gamma rays.
Galactic plane (GP) gamma rays are produced either by astrophysical sources or by cosmic rays interacting with the interstellar medium. Both processes have hadronic and leptonic contributions~\cite{Fermi-LAT:2012edv}.
Hadronic interactions produce neutral and charged pions. Neutral pions decay into gamma rays; the decay of charged pions results in neutrinos, which after oscillations are observed at Earth with a flavor ratio of approximately 1:1:1~\cite{Ahlers:2013xia}.

Efforts to detect neutrinos from the GP have been undertaken for many years 
by the IceCube, ANTARES, and Baikal-GVD neutrino telescopes \cite{IceCube:2017trr,ANTARES:2018nyb,ANTARES:2022izu,antaresTemplateSearch,Baikal-GVD:2024kfx}. 
The first significant evidence came in 2023, when
IceCube reported neutrino emission from the GP at \num{4.5}$\sigma$~\cite{IceCube:2023ame}.
This result has since been updated, reaching \num{5.7}$\sigma$~\cite{abbasi2026highenergyneutrinoemissionmilky}.
These searches are optimized to determine if the Milky Way as a whole is a neutrino source. 
Spatial and energy templates, based on model predictions that used GeV--TeV~\cite{Fermi-LAT:2012edv, Gaggero:2015xza} gamma-ray observations, were used to search for neutrinos.
Given that neutrino emission from the Galaxy is now firmly established, we develop a different approach: 
we present a characterization of the neutrino energy spectrum in various parts of the GP (segments). 
In each segment, we describe the spectrum assuming a power law in energy, and we fit both the flux normalization 
and the spectral index, independent of neutrino emission models.

The diffuse GeV gamma-ray emission of the GP has been measured by the space-based \textit{Fermi} Large Area Telescope (LAT)~\cite{Fermi-LAT:2012edv}.
Ground-based observatories have measured diffuse Galactic gamma-ray emission in the \unit{\tera\electronvolt} range, with
imaging atmospheric Cherenkov telescopes, such as H.E.S.S.~\cite{HESS:2016pst} and air-shower arrays 
such as HAWC~\cite{Alfaro_2024}, Tibet~AS$\gamma$~\cite{TibetASgamma:2021tpz}, and LHAASO~\cite{LHAASO:2023gne,LHAASO:2024lnz}.
Many TeV gamma-ray sources in the GP have been found~\cite{HESS:2018pbp}.
Galactic neutrino sources have not been detected~\cite{IceCube:2025lev}, and the neutrino flux reported here includes both diffuse GP and source emission.
The irreducible background for neutrino-source searches consists of atmospheric neutrinos
and the isotropic component of the astrophysical neutrino flux.
Their combined flux, measured by IceCube~\cite{IceCube:2025tgp,Abbasi:2025rmj}, far exceeds that expected from any single source, making individual source detection challenging.

There is a variety of models for diffuse emission from the GP. They simulate cosmic-ray transport throughout the Galaxy
so they reproduce the measured diffuse gamma-ray emission.
They then utilize the aforementioned connection between hadronic gamma rays and their neutrino 
counterpart to predict the spatial and energy distribution of the neutrino emission.
The earlier $\pi^0$ model extrapolated a power law with spectral index $\gamma=2.7$ 
from GeV energies~\cite{Fermi-LAT:2012edv} and the $\mathrm{KRA}_\gamma$ model~\cite{Gaggero:2015xza}
proposed radially dependent diffusion coefficients in the Galaxy, resulting in a harder spectral index
towards the Galactic center. 
Newer models account for gamma-ray measurements at higher energies, such as the updated $\mathrm{KRA}_\gamma$ models~\cite{DeLaTorreLuque:2025zsv} (called min and max) or 
the GALPROP model~\cite{marinos_simulating_2026}. 
The CRINGE model~\cite{Schwefer:2022zly} is fitted to recent Galactic cosmic-ray data, while also incorporating a model of 
unresolved sources, based on the pulsar-powered sources following~\cite{vecchiotti_cringe_2022}.

\section{Detector \& Data Sample}
The IceCube observatory with its upgrade deployed in 2026 consists of 5665 optical modules, each equipped with photomultiplier tubes, 
frozen inside 91 holes in a volume of one cubic \unit{\kilo\meter} of glacial ice \qty{1.5}{\kilo\meter} below the geographic South Pole~\cite{Aartsen:2016nxy,IceCubeCollaborationP:2025rpl,Eimer:2026qrp}.
The data used in this analysis is from the IceCube detector configuration without the recently deployed upgrade,
and is from 5160 optical modules on 86 strings.

When a high-energy neutrino interacts with the ice, it produces charged secondary particles, which 
emit Cherenkov light. 
The Cherenkov light can be used to reconstruct the direction and energy of the primary neutrino.
IceCube records two main event morphologies:
\enquote{tracks} and \enquote{cascades}. The former consists of Cherenkov light emitted along paths of muons
created through charged-current interactions of muon neutrinos.
Tracks provide the best angular resolution ($<\qty{1}{\degree}$ above $E_\nu=\qty{1}{\tera\electronvolt}$ \cite{IceCube:2025gdd}), 
but have a down-going (southern sky) background of atmospheric muons, which is $\sim 10^6$ times more common than any neutrino interaction.
\enquote{Cascades} occur when any flavor of neutrino interacts inside the detector volume via a
$Z^0$ boson exchange, or when electron neutrinos $\nu_e$ interact via any interaction type.
The majority of tau neutrinos $\nu_\tau$ also produce cascades.
The light from these cascades is emitted almost spherically and the angular resolution is thus poorer ($\sim \qty{7}{\degree}$ \cite{IceCube:2023ame,IceCube:2024csv}).
As down-going muons are not a main background, cascades are sensitive over the entire sky.
This work uses the same machine-learning-based cascade event selection and reconstruction~\cite{Huennefeld:2021Jk,Abbasi_2021} as the initial GP neutrino observation by IceCube~\cite{IceCube:2023ame}. 
The dataset comprises \num{59592} events recorded between May 2011 and May 2021
ranging from \qty{500}{\giga\electronvolt} to the \unit{\peta\electronvolt} region.

\section{Segmenting the Signal Hypothesis}
This work uses the same unbinned maximum-likelihood technique~\cite{Braun:2008bg} as previous IceCube 
neutrino source searches \cite{IceCube:2019cia, IceCube:2023ame}:
The signal probability distribution function (PDF) $S_i$ for each event $i$
is derived from Monte Carlo (MC) simulations, while the background PDF $B_i$ is constructed from data
by assuming a uniform distribution of background in right ascension while subtracting the signal contribution~\cite{IceCube:2017trr,IceCube:2023ame}.
Unlike previous neutrino source searches, we introduce multiple signal components in the likelihood and fit them simultaneously.
In this work, each component is a segment (labeled $k$) of the GP, whose flux $\Phi_k(E)$ is described 
by a single power law $\Phi_k(E_\nu) = \Phi_{0,k} E_\nu^{-\gamma_k}$ emitted with a uniform spatial distribution within the segment.
The spectral index $\gamma_k$ and flux normalization $\Phi_{0,k}$ are fitted to the data by maximizing the likelihood.

The bounds of the segments and their number can be freely chosen. 
We refer to this choice as a \enquote{segmentation scheme}.
In this work, we consider spatially disjoint segments aligning with the GP, so that 
the whole segmentation scheme fills the GP entirely, as shown in \Cref{fig:generic-3}.
Detailed information on the new method is available in the Supplemental Material~\cite{SM}.
This analysis enables a measurement of the spectral index of large-scale emission and of its variation across the sky.

In total, eight segmentation schemes are used to analyze the data. 
The first three are \enquote{generic} schemes, designed to probe how the flux and spectrum vary with Galactic longitude at increasingly finer resolution (two, three, and six segments). All three span Galactic latitudes from \qty{-8}{\degree} to \qty{8}{\degree}.
This latitude range aligns with the inner Galactic analysis region of \textit{Fermi}-LAT in~\cite{Fermi-LAT:2012edv},
and is also well suited to this analysis.
Since typical signal events of this data sample have a median angular uncertainty of \qty{7}{\degree}~\cite{IceCube:2023ame},
a latitude band narrower than $\sim \qty{14}{\degree}$ does not improve the sensitivity of this analysis.
The exact ranges of the generic segments are shown in \Cref{tab:generic}.

To enable direct comparison of neutrino emission with diffuse gamma-ray emission,
we analyze six regions in five segmentation schemes aligning with analysis regions of high-energy gamma-ray observatories:
HAWC~\cite{Alfaro_2024}, H.E.S.S.~\cite{HESS:2018pbp}, two regions from LHAASO~\cite{LHAASO:2023gne, LHAASO:2024lnz} 
and two from Tibet~AS$\gamma$~\cite{TibetASgamma:2021tpz}.
An overview of all gamma-ray aligned regions can be found in \Cref{tab:experiments} and a visualization of 
every fitted segment is shown in \Cref{fig:overview}.

\section{Results}
\begin{figure}[h]
\includegraphics[width=\linewidth]{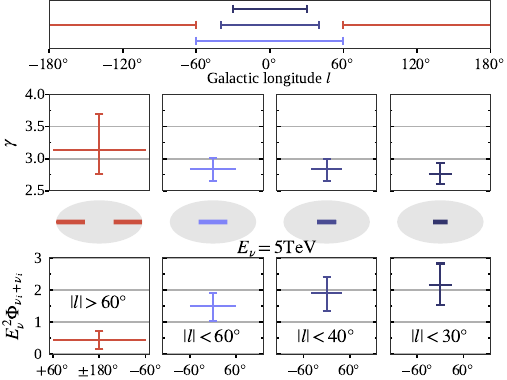}
\caption{Fitted spectral indices $\gamma$ (middle row) and flux normalizations $E^2 \Phi$ at $E_\nu=\qty{5}{\tera\electronvolt}$
in units of $10^{-10}\,\unit{\tera\electronvolt\centi\meter\tothe{-2}\second\tothe{-1}\steradian\tothe{-1}}$ (bottom row) 
of neutrino emission in the outer Galaxy (red) and different inner Galaxy regions (blue).
The regions are different in Galactic longitude, but the Galactic latitude range stays constant with $|b| < \qty{8}{\degree}$.  They are visualized in the top row, showing the overlap of the three different inner Galaxy regions, which are correlated.
The four regions in equatorial coordinates are also visualized between the lower two rows.
The leftmost (dark red) bar refers to the outer Galaxy at $\vert l\vert > \qty{60}{\degree}$.  
The three panels on the right show increasingly smaller regions of the inner Galaxy. 
}
\label{fig:hardening}
\end{figure}

All fitted values for the generic segmentation schemes are listed in \Cref{tab:generic}.
At a reference energy of \qty{5}{\tera\electronvolt}, the measured neutrino flux per solid angle
in units of $10^{-11}\,\unit{\tera\electronvolt\centi\meter\tothe{-2}\second\tothe{-1}\steradian\tothe{-1}}$
from the inner Galaxy $|l| < \qty{30}{\degree}$ is
$\num{21.4(68:61)}$, which is approximately
a factor of 5 higher than
$\num{4.3(28:28)}$, 
the measurement from the outer Galaxy $|l| > \qty{60}{\degree}$.
\Cref{fig:hardening} compares the measurements of the outer Galaxy with progressively smaller regions of the inner Galaxy.
The three regions around the Galactic center are increasingly small ($|l| < \qty{60}{\degree}, \qty{40}{\degree}, \qty{30}{\degree}$)
and overlap, meaning the results are correlated.
The single-parameter 68\% confidence intervals shown are obtained with the profile-likelihood method,
utilizing a Feldman--Cousins construction~\cite{PhysRevD.57.3873}, while 
also incorporating systematic uncertainties~\cite{Cousins:1991qz, Conrad:2002kn, Hill:2003jk}.
We find that after accounting for known sources of systematic uncertainties,
statistical uncertainties still dominate (for details see Supplemental Material~\cite{SM}).

The middle row of \Cref{fig:hardening} shows the fitted spectral indices 
for the different tested Galactic regions.
The innermost Galactic region is measured with a spectral index of 
$\gamma = \num{2.76(18:15)}$, 
improving on the previous result by ANTARES for the same Galactic longitude region~\cite{ANTARES:2022izu}.
A detailed comparison of the two results, which are compatible within uncertainties, can be found in the \nameref{sec:EndMatter}.
The outer Galaxy is fitted with a spectral index of 
$\gamma = \num{3.13(57:37)}$. This is softer than the 
fit for the inner Galaxy, consistent with expectation of~\cite{Gaggero:2015xza},
although, given the current uncertainties, the two indices remain compatible.

The \qty{5}{\tera\electronvolt} reference energy in \Cref{fig:hardening} was chosen as a compromise between the inner and outer Galaxy so that it is close to the central energy range driving the fits.
The energy range is evaluated with the same data-based method as~\cite{IceCube:2023ame}, and is described in the Supplemental Material~\cite{SM}.
As shown in \Cref{fig:hardening}, at \qty{5}{\tera\electronvolt} the inner 
Galaxy ($|l|<\qty{30}{\degree}$) emits an approximately five times higher neutrino flux per solid angle than the outer Galaxy ($|l|>\qty{60}{\degree}$).
The profiled confidence intervals are approximately 2.5 standard deviations apart at that energy.

\begin{figure}
\includegraphics[width=\linewidth]{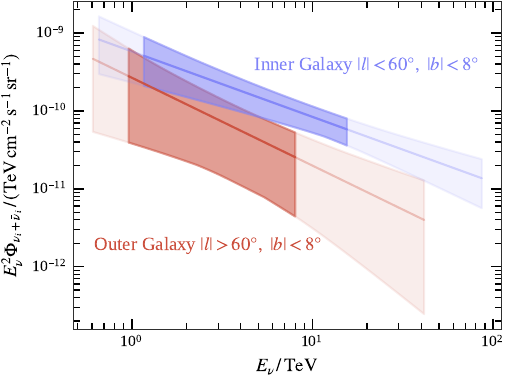}
\caption{Measurement of the average per-flavor neutrino 
energy flux $E^2\Phi(E)$ in the inner 
(blue, $|l| < \qty{60}{\degree}$, $|b| < \qty{8}{\degree}$) and 
outer (red, $|l| > \qty{60}{\degree}$, $|b| < \qty{8}{\degree}$) Galaxy for the 2-segment scheme. 
Measurements are given per solid angle to account for the different size of the two regions. 
The solid lines denote the best-fit flux and the 
enclosing regions the 68\% contour.
The dark (light) shaded regions refer to the 68\% (95\%) sensitive 
energy ranges.}
\label{fig:InnerOuterButterflies}
\end{figure}

A comparison of the neutrino emission across energies in different sky regions 
is shown in \Cref{fig:InnerOuterButterflies}.
Unlike the profiled intervals in
\Cref{fig:hardening}, the bands here represent the 2D 68\% uncertainty region in a $\gamma$-flux diagram.
The inner segment shows a higher emission than the outer one, though the exact ratio depends on the reference energy because the fitted spectral indices differ. 
The harder inner-Galaxy spectrum also shifts the energy range to higher energies than for the softer outer segment.

\begin{figure}
\includegraphics[width=\linewidth]{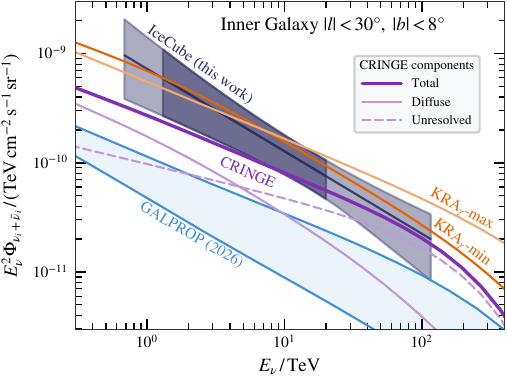}
\caption{Comparison of the fitted flux of the inner Galaxy in this work (68\% contour)
and three model predictions (CRINGE~\cite{Schwefer:2022zly}, the updated KRA$_{\gamma}$ models~\cite{DeLaTorreLuque:2025zsv} and  
the newly released GALPROP model~\cite{marinos_simulating_2026}). 
All fluxes refer to the inner Galaxy ($|l| < \qty{30}{\degree}, |b| < \qty{8}{\degree}$). The diffuse and unresolved sources~\cite{vecchiotti_cringe_2022} components 
of the CRINGE model are shown individually as well.
The dark (light) shaded region depicts the 68\% (95\%) sensitive energy range.}
\label{fig:ModelComparison}
\end{figure}

In \Cref{fig:ModelComparison}, we compare our inner-Galaxy measurement ($|l|<\qty{30}{\degree}$) 
with the predictions of the CRINGE~\cite{Schwefer:2022zly}, the KRA$_{\gamma}^{\min/\max}$~\cite{DeLaTorreLuque:2025zsv} and the GALPROP~\cite{marinos_simulating_2026} models.
Both KRA$_{\gamma}^{\min/\max}$ predictions are consistent with our result within uncertainties, 
while the flux predicted by the GALPROP model is a factor of $\sim$3--6 lower than
our best fit in the 68\% energy range.
The diffuse component of CRINGE underpredicts the flux in the inner Galaxy, while its total prediction including unresolved sources comes closer to the data.
The fluxes obtained in this work include all excess neutrino emission above the isotropic background flux, 
and consequently also include any unresolved source component on top of the diffuse emission.
The updated KRA$_\gamma$ models and the GALPROP model do not include any neutrino emission from sources.

The segmentation scheme with the highest significance, $3.84\sigma$, is the 3-segment scheme. 
The significance of this analysis is $\sim 1\sigma$ lower compared to the template analysis on the same dataset~\cite{IceCube:2023ame}.
As detailed in the \nameref{sec:EndMatter}, this is expected from pseudo-experiments, partially due to the larger number of free parameters. 

\section{Comparison to Gamma-ray Observations}
Fitting the neutrino flux normalization and spectral index independently in different segments
of the GP enables direct comparison with gamma-ray measurements. 
This was not possible previously using the all-sky template analysis, 
since the relative flux from different regions is fixed by the template itself, and only the normalization of the Galaxy as a whole 
is a free parameter.
To derive the neutrino-equivalent fluxes from the gamma-ray fluxes, we follow~\cite{Fang:2023azx,Ahlers:2013xia} and assume a
purely hadronic pion-production 
scenario in $pp$ collisions of cosmic rays with the interstellar medium, in which the $\pi^+\!:\!\pi^-\!:\!\pi^0$ production
ratio is 1:1:1.
Assuming Galactic gamma rays reach Earth unattenuated, the 
neutrino-equivalent fluxes can be obtained by
$E_\nu^2 \frac{\mathrm{d}N_{\nu_i}}{\mathrm{d}E_\nu} = \sfrac{1}{2} E_\gamma^2 \frac{\mathrm{d}N_{\gamma}}{\mathrm{d}E_\gamma}$ and $E_\nu = \sfrac{1}{2} E_\gamma$.

\begin{figure}
\includegraphics[width=\linewidth]{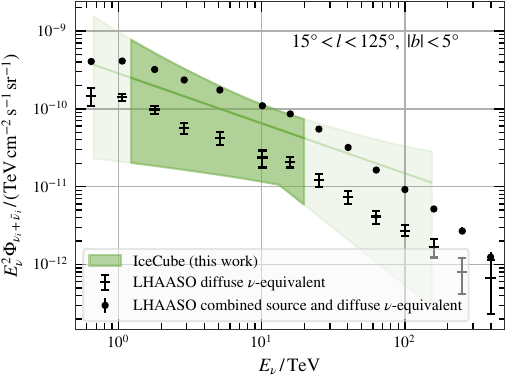}
\caption{The TeV gamma-ray flux reported by LHAASO from the GP is converted into a neutrino-equivalent flux
and compared against our measurement of the exact same region assuming a power-law spectrum.
We show both LHAASO's diffuse measurement obtained by masking known gamma-ray sources from the region (black crosses) \cite{LHAASO:2024lnz}
and LHAASO's unmasked measurement of the region including all gamma-ray sources (black dots)~\cite{LHAASO:2024lnz:suppl}.
The dark shaded region refers to the 68\% contour of our power-law fit with the 68\% and 95\% sensitive 
energy ranges provided, while the dark green line is our best-fit flux.
}\label{fig:lhaaso_comparison}
\end{figure}

We focus on the comparison to LHAASO, whose
measurement of the diffuse emission spans \qty{1}{\TeV} to
\qty{1}{\peta\electronvolt}~\cite{LHAASO:2023gne,LHAASO:2024lnz}, covering our entire energy range.
Best-fit values and confidence intervals for the regions aligned
with HAWC, H.E.S.S., and Tibet~AS$\gamma$ are listed in \Cref{tab:experiments}.
In \Cref{fig:lhaaso_comparison}, we show a comparison of our
neutrino measurement to the gamma-ray measurements by LHAASO using 
the conversion described above.
We show both LHAASO's diffuse-only estimate, obtained by masking known sources~\cite{LHAASO:2024lnz}, and its unmasked total flux~\cite{LHAASO:2024lnz:suppl} in the region $\qty{15}{\degree} < l < \qty{125}{\degree}$.
LHAASO's data points for the total flux in neutrino-equivalent form fall above our best fit and within our 68\% uncertainty contour, which lets us 
conclude compatibility since the hadronic fraction must be one or less.
The diffuse estimate of LHAASO falls below our best fit and within uncertainties,
while their fitted spectral index of $\gamma=2.66\pm0.05_{\mathrm{stat}}$ before the reported break at $E_\gamma=32.84\pm11.16_{\mathrm{stat}}\unit{\TeV}$
is compatible with our fitted spectral index for the region of $\gamma=2.64^{+0.41}_{-0.40}$.
Since both LHAASO's diffuse-only and total TeV Galactic fluxes are 
compatible with our measurement,
leptonic emission scenarios for the gamma-ray sources cannot be ruled out. 

\section{Conclusion}
In this work,  we report measurements of the neutrino emission along the GP.
We introduce a likelihood method that allows simultaneous measurements in different regions along the whole plane without assumptions about the relative emission in each region. 
We report the measured power-law spectral index and flux normalization of the neutrino emission in each studied region. 
The best-fit spectral index is harder in the inner Galaxy ($\abs{l} < \qty{30}{\degree}$) compared to the outer Galaxy ($\abs{l} > \qty{60}{\degree}$) but is 
similar within uncertainties,
while the flux normalization in the inner Galaxy is approximately a factor of 5 higher than in the outer Galaxy at $E_\nu=\qty{5}{\tera\electronvolt}$.
The measurement presented in this work for the inner Galaxy is the most precise to date, compatible with the previously reported measurement 
by ANTARES while shrinking the uncertainty contour on flux and spectral index significantly. 

Comparing to model predictions, we find the diffuse-only component of the CRINGE model underpredicts the measured neutrino emission from the inner Galaxy,
and is brought closer to our measurement with the inclusion of predicted emission from 
unresolved sources.
Similarly, the GALPROP model would require a component of additional emission on top of its 
predicted diffuse neutrino emission to be compatible with our measurements.
The updated KRA${}_\gamma^{\mathrm{min/max}}$ models, on the other hand, are consistent with our neutrino measurements in the same region, taking into account diffuse emission only. 
We also present a measurement of 1--100\,\unit{\tera\electronvolt} neutrino emission in the region $\qty{15}{\degree}<l<\qty{125}{\degree}$ of the GP, where LHAASO has reported a measurement of the diffuse gamma-ray emission in the same energy range.  
Comparing to both the diffuse-only and the total gamma-ray flux measured by LHAASO in this region, we find a neutrino flux that is in between the two and compatible with both, assuming fully hadronic gamma-ray emission.

The results of this analysis demonstrate that it is becoming possible to test models of Galactic diffuse emission.
Combined with improved data samples from IceCube~\cite{abbasi2026highenergyneutrinoemissionmilky} and new data from 
neutrino telescopes in the northern hemisphere under construction now~\cite{aiello_astronomy_2024,KM3NeT:2026omo}, such direct comparisons between neutrino and gamma-ray measurements will become increasingly precise, and ultimately directly constrain the hadronic fraction of gamma-ray observations in the plane of the Galaxy.

\section{Acknowledgments}
The authors would like to thank Pedro De la Torre Luque for help and discussion in providing the $\mathrm{KRA}_{\gamma}^{\min/\max}$ models.
The authors would also like to thank Peter David Marinos for providing the GALPROP model ahead of publication.
The IceCube Collaboration acknowledges the significant contributions to this manuscript from Mirco Hünnefeld and Ludwig Neste.
\input{acknowledgments.tex}
\bibliography{references}

\FloatBarrier

\ifarxiv
    \clearpage
\else
    \onecolumngrid    
    \vspace{1.5em}    
    \twocolumngrid    
\fi

\appendix
\section{End Matter}\label{sec:EndMatter}

In \Cref{tab:experiments,tab:generic}, we present the results and confidence intervals for all fit parameters.
The profiled confidence intervals for each parameter include known systematic uncertainties stemming from the detector and ice modeling,
and are constructed by inversion of the likelihood ratio test~\cite{PhysRevD.57.3873, Cousins:1991qz, Conrad:2002kn, Hill:2003jk} (for a detailed explanation, see Supplemental Material~\cite{SM}). 
The significance of each segmentation scheme, consisting of one choice to divide the GP into different segments, is obtained by comparing the obtained test statistic against 
the empirical distribution of test statistics from simulated pseudo-experiments under the null hypothesis
of no neutrino emission, obtained by randomizing the right-ascension coordinates of the data events.
The highest significance of $3.84\sigma$ is obtained for the division of the GP into three segments.
Considering that eight different segmentation analyses were performed, the global significance (found by performing all eight analyses on scrambled datasets) is $3.43\sigma$. 
The significances for each segment on its own have been obtained post-unblinding by comparing the $\Delta \ln \mathcal{L} = \ln \mathcal{L}(\hat n_k, \hat \gamma_k) - \ln \mathcal{L}(n_k=0)$
for each segment $k$ to the distribution of the same value under the null hypothesis and keeping the parameters of all other segments fixed at their best-fit point. 

Since this work uses the same data sample as the earlier template fit~\cite{IceCube:2023ame},
we expect the significance of this work to be smaller.
The segmented analysis presented here cannot in general provide higher significance than the single template analysis, unless the assumed template differs substantially from the true emission.
To quantify this, we simulated a
neutrino flux according to the distribution of
the $\pi^0$ model at the best-fit normalization of 749 neutrinos~\cite{IceCube:2023ame}.
For each pseudo-experiment we performed both the template analysis assuming 
the $\pi^0$ model and the segmented analysis of this work, dividing 
up the GP into three segments with a total of six parameters.
The fixed-spectrum template analysis results in a median significance of 
$4.7\sigma$, while the segmented analysis results in a median significance of $3.2\sigma$. 

These can be compared to the significances of these analyses on the real data, which 
were 4.7$\sigma$ for the template-analysis and $3.84\sigma$ for the segmented analysis,
indicating that the assumed distribution and spectral index of the model
are accurate within uncertainties to fit our data.
Consistently, the fitted inner-Galaxy index of $\gamma =  \num{2.76(18:15)}$ agrees within uncertainties with the $\gamma = 2.7$ of the $\pi^0$ model.

\begin{figure}
\includegraphics[width=\linewidth]{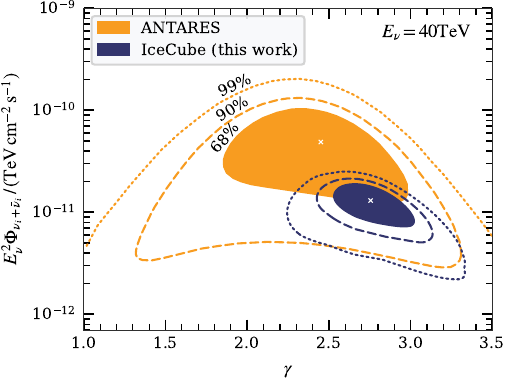}
\caption{Contours at $E_\nu=\qty{40}{\tera\electronvolt}$ of the solid-angle integrated flux normalization 
and spectral index $\gamma$, in the Galactic ridge $|l| < \qty{30}{\degree}$ of this work ($|b| < \qty{8}{\degree}$) 
and ANTARES ($|b| < \qty{2}{\degree}$)~\cite{ANTARES:2022izu}. The white crosses denote the best-fit point of each analysis.
The contours of this work assume Wilks' theorem~\cite{Wilks:1938dza}. }
\label{fig:antares}
\end{figure}

\section{Comparison to ANTARES}\label{sec:antares}
The ANTARES Collaboration presented a power-law fit and an on-off counting analysis of the Galactic ridge ($|l| < \qty{30}{\degree},\, |b| < \qty{2}{\degree}$)~\cite{ANTARES:2022izu}.
The power-law fit in this region found a $2.0\sigma$ excess, while the counting analysis 
resulted in $2.2\sigma$ significance.
Here, we compare the results of this work and ANTARES for the same Galactic longitude region.
The widths of the latitude regions differ, reflecting the different angular resolution of the primary contributing event type for 
each detector (down-going cascades for IceCube and primarily through-going muon track events for ANTARES).
The $|b| < \qty{8}{\degree}$ latitude range of this work is four times larger than 
the region chosen by ANTARES.
\Cref{fig:antares} compares the 68\%, 90\%, and 99\% contours in solid-angle integrated flux and spectral index of this work and the ANTARES result at $E_\nu=\qty{40}{\tera\electronvolt}$.
The reference energy has been adjusted from the previously reported $E_\nu=\qty{5}{\tera\electronvolt}$ to match the energy at which ANTARES is reporting the flux. 
To assess the impact of the different latitude ranges, we use model predictions:
At $E_\nu = \qty{40}{\TeV}$, the predicted fraction of the $|l|<30^\circ$, $|b|<8^\circ$ flux that falls within $|b|<2^\circ$
is \qty{86}{\percent} for KRA$_\gamma^{\mathrm{min/max}}$, \qty{83}{\percent} for KRA$_\gamma^{\mathrm{5 PeV/50 PeV}}$, \qty{61}{\percent} for CRINGE, \qty{66}{\percent} for GALPROP, and \qty{65}{\percent} for $\pi^0$.
The 99\% contour of this work is mostly contained within the ANTARES 90\% contour, 
which implies agreement under the assumption that the neutrino flux in the region $\qty{2}{\degree} < |b| < \qty{8}\degree$ can be neglected.
However, this new result suggests that the neutrino 
spectral index is softer than the ANTARES best-fit value and---keeping in mind 
the different latitude regions---the flux is smaller than reported by ANTARES.
This is consistent with a recent re-analysis of the full ANTARES data~\cite{antaresTemplateSearch},
which did not find Galactic emission beyond 1.3$\sigma$ with the template method and, 
when repeating the counting analysis of the Galactic ridge with 2 more years of data (corresponding to about a \qty{15}{\percent} increase in data),
found a lower significance of neutrino emission in the Galactic ridge (1.9$\sigma$~\cite{antaresTemplateSearch} compared to the earlier 2.2$\sigma$ result~\cite{ANTARES:2022izu}).
This would require the estimated signal strength of ANTARES in the Galactic ridge to be lowered in light of new data.

\begin{table*}[ht]
    \centering
    \caption{Overview of the generic segmentation schemes. $l_{\mathrm{min}/\mathrm{max}}$ is the minimal/maximal Galactic longitude of the given segment. 
    All segments are in between \qty{\pm 8}{\degree} of Galactic latitude $b$. 
    The summed number of signal neutrinos $n_s$ from all segments and the significance of the analysis are also given,
    as well as the best-fit spectral indices and best-fit fluxes $E^2\Phi_{\nu_i+\bar\nu_i}$ in $\qty{1e-11}{\tera\electronvolt\centi\meter\tothe{-2}\second\tothe{-1}\steradian\tothe{-1}}$ at \qty{5}{\tera\electronvolt}.
    All intervals are 68\% confidence intervals.
    }\label{tab:generic}
    \begin{tblr}{
        colspec={l Q[si={table-format=+3.0}] Q[si={table-format=+3.0}] Q[si={table-format=1.2}] Q[si={table-format=3.0}] Q[si={table-format=2.4}] Q[si={table-format=1.6}] },
        row{1}={guard,c},
        cell{2,5,9}{1} = {c=7}{c,guard},
        cell{3-4,6-8,10-15}{2-3} = {appto={\unit{\degree}}},
        cell{3-4,6-8,10-15}{4} = {appto={$\sigma$}},
    }
      \toprule
      Name & $l_{\mathrm{min}}$ & $l_{\mathrm{max}}$ & Significance & $n_k$ & $E^2\Phi_{\qty{5}{\tera\electronvolt}}$ & $\gamma$  \\
      \midrule
        2 Segments (Total: $3.81\sigma$, $n_s=645$) \\
        \midrule
        $S_1$/inner & -60 & 60   & 3.83 & 330 & 15.0(39:46) & 2.84(17:18) \\
        $S_2$/outer & 60 & -60   & 1.28 & 315 & 4.3(28:28) & 3.13(57:37) \\
      \midrule
        3 Segments  (Total: $3.84\sigma$, $n_s=643$) \\
      \midrule
        $S_1$/inner & -40 & 40   & 3.75 & 269 & 19.1(50:56) & 2.83(15:18) \\
        $S_2$/right & 40 & 180   & 1.27 & 243 & 5.1(39:39) & 2.93(81:71) \\
        $S_3$/left & -180 & -40 & 0.71 & 130 & 3.5(44:28) & {{{$>$ 2.87}}} \\
      \midrule
        6 Segments  (Total: $3.58\sigma$, $n_s=816$) \\
      \midrule
        $S_1$/inner & -30  & 30   & 3.35 & 223 & 21.4(68:61) & 2.76(18:15) \\
        $S_2$ & 30 & 90    & 1.84 & 252 & 11.6(61:78) & 3.32(52:42) \\
        $S_3$ & 90 & 150   & 1.12 & 20 & 0.1(39:1) & 1.77(34:21) \\
        $S_4$ & 150 & -150 & 1.63 & 235 & 9.0(60:59) & 3.41(50:28) \\
        $S_5$ & -150 & -90 & 0.58 & 51 & {{{$<$10.84}}} & {{{-}}} \\
        $S_6$ & -90 & -30  & 0.20 & 36 & {{{$<$7.96}}} & {{{-}}} \\
      \bottomrule
    \end{tblr}
\end{table*}

\begin{table*}[ht]
    \centering
    \caption{Overview of the five segmentation schemes used to align with the Galactic analysis regions of gamma-ray telescopes. 
    For each scheme, the remainder of the GP at the same latitude extent is also fitted and labeled \enquote{outside}. The LHAASO scheme contains two gamma-ray-aligned regions, listed separately.
    The significance of the scheme is given, as well as the sum of the best-fit number of signal neutrinos from all segments $n_s$.
    The results for the best-fit fluxes $E^2\Phi_{\nu_i+\bar\nu_i}$ are given in $\qty{1e-11}{\tera\electronvolt\centi\meter\tothe{-2}\second\tothe{-1}\steradian\tothe{-1}}$ at \qty{5}{\tera\electronvolt}.
    All intervals are 68\% confidence intervals.
    }
    \label{tab:experiments}
   \begin{tblr}{
        colspec={l Q[si={table-format=+3.0}]  Q[si={table-format=+3.0}] Q[si={table-format=1.0}] Q[si={table-format=1.2}] Q[si={table-format=3.0}] Q[si={table-format=1.4}] Q[si={table-format=1.6}]},
        row{1}={guard},
        cell{3-4,7,10,13,16,19}{2-3} = {appto={\unit{\degree}}},
        cell{3-5,7-8,10-11,13-14,16-17,19-20}{4} = {appto={\unit{\degree}}},
        cell{3-5,7-8,10-11,13-14,16-17,19-20}{5} = {appto={$\sigma$}},
    }
    \toprule
    Name & $l_{\mathrm{min}}$ & $l_{\mathrm{max}}$ & $ \lvert{b}\rvert_{\mathrm{max}}$ & Significance & $n_k$ & $E^2\Phi_{\qty{5}{\tera\electronvolt}}$ & $\gamma$  \\
    \midrule
    \SetCell[c=8]{c,guard} LHAASO \cite{LHAASO:2023gne,LHAASO:2024lnz} (Total: $2.92\sigma, n_s=567$) \\
    \midrule
    $S_1$ & 15         & 125             & 5   & 1.80 & 177 & 10.1(63:71) & 2.64(41:40) \\
    $S_2$ & 125        & -125             & 5   & 0.79 & 171 & 4.3(80:31) & {{{$>$ 3.00}}} \\
    $S_3$ & \SetCell[c=2]{c,guard} outside & & 5   & 2.82 & 219 & 13.7(57:57) & 2.79(23:23) \\
    \midrule
    \SetCell[c=8]{c,guard} HAWC \cite{Alfaro_2024} (Total: $3.01\sigma, n_s=668$) \\
    \midrule
     $S_1$ & 43         & 73              & 5 &  0.87 & 122 & 13.5(196:101) & {{{$> 2.78$}}} \\
     $S_2$ & \SetCell[c=2]{c,guard} outside & & 5    & 2.97 & 546 & 10.8(37:35) & 2.85(20:19) \\
    \midrule
    \SetCell[c=8]{c,guard} H.E.S.S. \cite{HESS:2018pbp} (Total: $3.06\sigma, n_s=569$) \\
    \midrule
    $S_1$ & -110        & 65              & 3 & 2.89 & 310 & 24.1(89:83) & 2.88(24:21) \\
    $S_2$ & \SetCell[c=2]{c,guard} outside& & 3 & 1.23 & 259 & 11.9(82:76) & 3.03(60:40) \\
    \midrule
    \SetCell[c=8]{c,guard} Tibet~AS$\gamma$ I \cite{TibetASgamma:2021tpz} (Total: $2.96\sigma, n_s=660$) \\
    \midrule
    $S_1$ & 25         & 100              & 5 & 1.88 & 217 & 16.7(83:89) & 2.98(51:39) \\
    $S_2$ & \SetCell[c=2]{c,guard} outside & & 5 & 2.60 & 443 & 10.4(39:39) & 2.88(22:22) \\
    \midrule
    \SetCell[c=8]{c,guard} Tibet~AS$\gamma$ II \cite{TibetASgamma:2021tpz} (Total: $2.97\sigma, n_s=626$) \\
    \midrule
    $S_1$ & 50         & -160        & 5         & 1.33 & 295 & 8.2(58:50) & 3.13(53:46) \\
    $S_2$ & \SetCell[c=2]{c,guard} outside & & 5   & 2.81 & 331 & 12.8(56:45) & 2.86(21:22) \\
    \bottomrule
    \end{tblr} 
\end{table*}

\clearpage
\onecolumngrid   
\phantomsection\label{sm:start}   

\setcounter{page}{1}
\renewcommand{\thepage}{S\arabic{page}}
\setcounter{figure}{0}
\renewcommand{\thefigure}{S\arabic{figure}}
\setcounter{table}{0}
\renewcommand{\thetable}{S\arabic{table}}
\setcounter{equation}{0}
\renewcommand{\theequation}{S\arabic{equation}}
\setcounter{section}{0}
\renewcommand{\thesection}{S\Roman{section}}

\renewcommand{\theHfigure}{S\arabic{figure}}
\renewcommand{\theHtable}{S\arabic{table}}
\renewcommand{\theHequation}{S\arabic{equation}}
\renewcommand{\theHsection}{S\Roman{section}}

\begin{center}
  {{\large\bfseries Supplemental Material for}\\[2pt]
  \enquote{\papertitle}}\\[8pt]
  \textit{The IceCube Collaboration}
\end{center}
\vspace{1em}

\FloatBarrier

\twocolumngrid

\section*{Likelihood Method}\label{sec:likelihood_in_depth}

The likelihood is a product over the $N$ observed events,
\begin{equation}
\mathcal{L} ({n_s}, {\gamma})=
\prod_{i=1}^{N} \left[
\frac{n_s}{N} S_i(E_i, \delta_i, \alpha_i, \sigma_i, \gamma)
+ \frac{N-{n_s}}{N} B_i(E_i, \delta_i)
\right],
\label{eq:likelihood_base}
\end{equation}
where for each event $i$ the reconstructed quantities are energy 
$E_i$,  declination $\delta_i$, right ascension $\alpha_i$ and angular uncertainty $\sigma_i$.
The number of signal neutrinos $n_s$ and the spectral index of 
the signal PDF $\gamma$ are optimized to maximize the likelihood of the data.
The signal PDFs $S_i$ for each event are constructed using Monte Carlo (MC) simulation, while the 
background PDFs $B_i$ are derived in a data-driven way.

Unlike previous neutrino searches, we introduce additional free parameters into the likelihood by segmenting
the signal PDF $S_i$ into $M$ parts
\begin{equation}
S_i 
= 
\sum_{k=1}^M
w_k S_i^k(E_i, \delta_i, \alpha_i, \sigma_i, \gamma_k).
\label{eq:segmented_sigpdf}
\end{equation}
Other than the total number of signal neutrinos $n_s$ and the 
spectral index $\gamma_k$ of each segment, we introduce the additional fit parameters 
$w_1,\dots, w_M$ as the fractional signal contribution of each segment $k$.
In order to make $S_i$ a PDF when segmenting, the weights must be normalized $\left(\sum_k w_k = 1\right)$ and positive $\left(w_k \geq 0\right)$, which is 
implemented via a softmax-transformation $w_k = e^{v_k}/\sum_{k'}e^{v_{k'}}$ in the fitting routine. With $n_k=w_k\cdot n_s$, the weights correspond to the number of signal neutrinos $n_k$ from segment $k$. 
The flux is calculated by transforming $n_k$ into the flux normalization $\Phi_0$ at a chosen reference energy $E_0$ with 
the detector's acceptance to the signal hypothesis of that segment $S_k$.

To account for the angular resolution and detector acceptance, the PDFs incorporate 
per-event detector acceptance and a convolution with the
point spread function.
This approach fits different energy distributions in different regions of the 
sky.

In this work, we consider spatially disjoint, uniform-emission PDFs, aligning with the GP in longitude.
While the joint likelihood accounts for correlations between segments, these turn out to be negligible on the unblinded data, which was 
expected from MC simulations.
We note that the segments in general do not need to be disjoint, 
and the formulation of the likelihood can be used to search for multiple different components 
of neutrino emission in the same region.
Because each segment's spatial PDF resembles uniform emission, 
an event's contribution to the likelihood ratio depends only on its angular uncertainty and the detector acceptance if it falls inside or near the segment; events far outside a segment do not contribute.
Compared to using realistic neutrino emission models as a template,  
the segmented method described here will yield a lower significance for finding neutrino emission from the GP.
Thus, the segmentation choices are not made to optimize the $p$-value;
template analyses \cite{IceCube:2023ame,abbasi2026highenergyneutrinoemissionmilky}
are more powerful for that purpose, provided the used models are close enough to the truth. 
Rather, the choice of the segmentation schemes is made to extract more physical parameters from the data,
while staying independent from concrete $\nu$-emission models.

In contrast to point-source searches, for extended-source searches like the GP the relative contribution of signal events to total events in the sample is not negligible,
and the background PDF cannot be constructed simply by averaging over right ascension at each declination.
Instead, the right-ascension-averaged signal PDF $\tilde S_i$ is subtracted from the averaged data PDF $\tilde D_i$ using 
the method described in \cite{IceCube:2017trr}, which was also used in the first GP observation analysis~\cite{IceCube:2023ame}.
This gives the \enquote{signal-subtracted} background PDF 
\begin{equation}
B_i^{\mathrm{S}}(E_i, \delta_i, \gamma) = \frac{N}{N-n_s} \left(\tilde D(\delta_i, E_i) - \frac{n_s}{N} \tilde S_i(\delta_i, E_i, \sigma_i, \gamma)\right).
\label{eq:sigsub}
\end{equation}
Substituting the last term in \Cref{eq:likelihood_base} with \Cref{eq:sigsub} results in the 
likelihood used in this work.

\begin{figure}
\includegraphics[width=\linewidth]{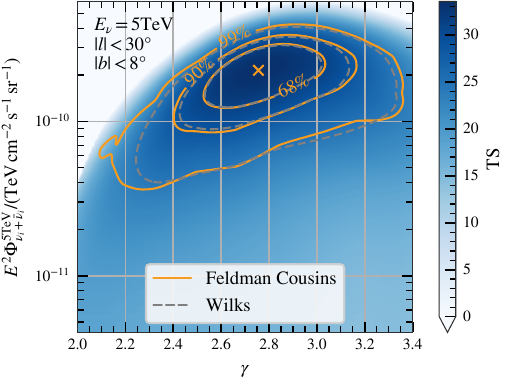}
\caption{The likelihood distribution as a function of $\gamma$ and the flux normalization at $E_\nu=\qty{5}{\tera\electronvolt}$.
The solid, orange contours are produced with the Feldman--Cousins procedure~\cite{PhysRevD.57.3873}, incorporating 
known sources of systematic uncertainties~\cite{Cousins:1991qz,Conrad:2002kn,Hill:2003jk}, while the grey, dashed contours are drawn assuming 
Wilks' theorem.
}\label{fig:llh_landscape}
\end{figure}

Using Eq. \eqref{eq:segmented_sigpdf} to describe a signal hypothesis with multiple source segments, the likelihood Eq. \eqref{eq:likelihood_base}
can be written as 
\begin{multline}
\mathcal{L} (n_1,\dots,n_M, \gamma_1, \dots, \gamma_M)=\\
\prod_{i=1}^{N} \left[
\frac{1}{N}   
\sum_{k=1}^{M}  n_k S^k_i
+ \frac{N-{n_s}}{N} B_i^{\mathrm{S}}
\right. \left.\vphantom{\sum_{k}^M}\right].
\label{eq:likelihood_segmented}
\end{multline}
The total number of signal neutrinos $n_s$ can be obtained 
by summing over the contribution from each region $n_s=\sum_k n_k$.
The likelihood-ratio test statistic
\begin{equation}
  \mathrm{TS} =  \mathrm{TS}(n_k=\hat n_k, \gamma_k=\hat\gamma_k) 
  = 2 \ln \frac{\mathcal{L}(n_k=\hat n_k, \gamma_k = \hat\gamma_k)}{\mathcal{L}(n_s=0)}
\label{eq:ts}
\end{equation}
is defined by the ratio of the likelihood where all parameters $n_k, \gamma_k$ ($k=1,\dots, M)$ are optimized to maximize the likelihood,
compared to the null hypothesis of no neutrino emission $n_s=0$ from the GP.
This approach takes care of any possible correlation between the segments, 
for example due to events contributing to multiple segments or a large influence from the signal subtraction.
However, the likelihood function of the unblinded data is essentially uncorrelated between the chosen segments.
This behavior was expected from studies of Monte Carlo simulations performed before analyzing the real data.
Thus, although the presented fits in this work are up to 12-dimensional (\enquote{6-segments}),
they can be represented as six two-dimensional slices in a $\gamma$-$\Phi$ space, without loss of information.
One of these slices is shown in \Cref{fig:llh_landscape} for the inner Galaxy for the differential flux per solid angle at \qty{5}{\tera\electronvolt}, 
which is proportional to the observed number of signal neutrinos in the segment $n_k$.
The color corresponds to the test statistic as a function of the flux per solid angle $E_\nu^2 \Phi_k$ and 
the spectral index $\gamma$.
The displayed TS is maximized over the parameters of all other segments; 
this profiling gives the same value as fixing them at their global best fit, indicating that the segments are uncorrelated.
We point out that the TS does not approach zero towards zero flux in that segment (or equivalently $n_k=0$),
as the other segments still contribute to the test statistic.
The energy ranges in this Letter are evaluated with the same data-based method as \cite{IceCube:2023ame},
letting each signal-like event contribute to the estimated energy distribution proportionally to 
its contribution to the test statistic at the best-fit $\gamma_k$ and $\Phi_{0,k}$, while ignoring background-like events which contribute negatively 
to the log-likelihood ratio.

The difference of the statistic at a point $\mathrm{TS}(\Phi_k, \gamma_k)$
compared to the best-fit point $\Delta \mathrm{TS} = \mathrm{TS}-\mathrm{TS}(\Phi_k, \gamma_k)$
is $\chi^2$-distributed in the case that Wilks' theorem is applicable~\cite{Wilks:1938dza}, 
and can in this case be easily used to construct a confidence region.
If Wilks' theorem does not hold, for example because of low statistics, 
simplified likelihood models or if one of the parameters lies at a 
boundary, 
$\Delta \mathrm{TS}$ can still be used to construct confidence regions 
by employing the likelihood ratio ordering method proposed by 
Feldman and Cousins~\cite{PhysRevD.57.3873}.
There, each point $n_k, \gamma_k$ is included in the $\alpha$-level confidence region
if the observed $\Delta \mathrm{TS}$ value at that point $n_k, \gamma_k$
is smaller than the $\alpha$-quantile of the distribution 
of $\Delta \mathrm{TS}$ values at that point.
For complex likelihoods this needs to be evaluated numerically,
and here, we additionally use a distribution of different 
realizations of systematic uncertainties over a Bayesian prior
as described in~\cite{Cousins:1991qz,Conrad:2002kn, Hill:2003jk}. 
The confidence regions obtained by Wilks' theorem and 
this systematic-incorporated Feldman--Cousins method 
are both displayed in \Cref{fig:llh_landscape}.
Especially for the 68\% and 90\% confidence regions they are 
very similar, indicating that Wilks' theorem is applicable and 
the statistics dominate the uncertainties.
Additionally, the Feldman--Cousins method automatically transitions 
the confidence interval to an upper limit, should the observed flux 
not be high enough, avoiding the flip-flopping problem~\cite{PhysRevD.57.3873}.

\end{document}

%% file: authorlist.tex
\affiliation{III. Physikalisches Institut, RWTH Aachen University, D-52056 Aachen, Germany}
\affiliation{Department of Physics, University of Adelaide, Adelaide, 5005, Australia}
\affiliation{Dept. of Physics and Astronomy, University of Alaska Anchorage, 3211 Providence Dr., Anchorage, AK 99508, USA}
\affiliation{School of Physics and Center for Relativistic Astrophysics, Georgia Institute of Technology, Atlanta, GA 30332, USA}
\affiliation{Dept. of Physics, Southern University, Baton Rouge, LA 70813, USA}
\affiliation{Dept. of Physics, University of California, Berkeley, CA 94720, USA}
\affiliation{Lawrence Berkeley National Laboratory, Berkeley, CA 94720, USA}
\affiliation{Institut f{\"u}r Physik, Humboldt-Universit{\"a}t zu Berlin, D-12489 Berlin, Germany}
\affiliation{Fakult{\"a}t f{\"u}r Physik {\&} Astronomie, Ruhr-Universit{\"a}t Bochum, D-44780 Bochum, Germany}
\affiliation{Universit{\'e} Libre de Bruxelles, Science Faculty CP230, B-1050 Brussels, Belgium}
\affiliation{Vrije Universiteit Brussel (VUB), Dienst ELEM, B-1050 Brussels, Belgium}
\affiliation{Dept. of Physics, Simon Fraser University, Burnaby, BC V5A 1S6, Canada}
\affiliation{Department of Physics and Laboratory for Particle Physics and Cosmology, Harvard University, Cambridge, MA 02138, USA}
\affiliation{Dept. of Physics, Massachusetts Institute of Technology, Cambridge, MA 02139, USA}
\affiliation{Dept. of Physics and The International Center for Hadron Astrophysics, Chiba University, Chiba 263-8522, Japan}
\affiliation{Department of Physics, Loyola University Chicago, Chicago, IL 60660, USA}
\affiliation{Dept. of Physics and Astronomy, University of Canterbury, Private Bag 4800, Christchurch, New Zealand}
\affiliation{Dept. of Physics, University of Maryland, College Park, MD 20742, USA}
\affiliation{Dept. of Astronomy, Ohio State University, Columbus, OH 43210, USA}
\affiliation{Dept. of Physics and Center for Cosmology and Astro-Particle Physics, Ohio State University, Columbus, OH 43210, USA}
\affiliation{Niels Bohr Institute, University of Copenhagen, DK-2100 Copenhagen, Denmark}
\affiliation{Dept. of Physics, TU Dortmund University, D-44221 Dortmund, Germany}
\affiliation{Dept. of Physics and Astronomy, Michigan State University, East Lansing, MI 48824, USA}
\affiliation{Dept. of Physics, University of Alberta, Edmonton, Alberta, T6G 2E1, Canada}
\affiliation{Erlangen Centre for Astroparticle Physics, Friedrich-Alexander-Universit{\"a}t Erlangen-N{\"u}rnberg, D-91058 Erlangen, Germany}
\affiliation{Physik-department, Technische Universit{\"a}t M{\"u}nchen, D-85748 Garching, Germany}
\affiliation{D{\'e}partement de physique nucl{\'e}aire et corpusculaire, Universit{\'e} de Gen{\`e}ve, CH-1211 Gen{\`e}ve, Switzerland}
\affiliation{Dept. of Physics and Astronomy, University of Gent, B-9000 Gent, Belgium}
\affiliation{Dept. of Physics and Astronomy, University of California, Irvine, CA 92697, USA}
\affiliation{Karlsruhe Institute of Technology, Institute for Astroparticle Physics, D-76021 Karlsruhe, Germany}
\affiliation{Karlsruhe Institute of Technology, Institute of Experimental Particle Physics, D-76021 Karlsruhe, Germany}
\affiliation{Dept. of Physics, Engineering Physics, and Astronomy, Queen's University, Kingston, ON K7L 3N6, Canada}
\affiliation{Department of Physics {\&} Astronomy, University of Nevada, Las Vegas, NV 89154, USA}
\affiliation{Nevada Center for Astrophysics, University of Nevada, Las Vegas, NV 89154, USA}
\affiliation{Dept. of Physics and Astronomy, University of Kansas, Lawrence, KS 66045, USA}
\affiliation{UCLouvain, Centre for Cosmology, Particle Physics and Phenomenology, CP3, Chemin du Cyclotron 2, 1348 Louvain-la-Neuve, Belgium}
\affiliation{Department of Physics, Mercer University, Macon, GA 31207-0001, USA}
\affiliation{Dept. of Astronomy, University of Wisconsin{\textemdash}Madison, Madison, WI 53706, USA}
\affiliation{Dept. of Physics and Wisconsin IceCube Particle Astrophysics Center, University of Wisconsin{\textemdash}Madison, Madison, WI 53706, USA}
\affiliation{Institute of Physics, University of Mainz, Staudinger Weg 7, D-55099 Mainz, Germany}
\affiliation{Department of Physics, Marquette University, Milwaukee, WI 53201, USA}
\affiliation{Institut f{\"u}r Kernphysik, Universit{\"a}t M{\"u}nster, D-48149 M{\"u}nster, Germany}
\affiliation{Bartol Research Institute and Dept. of Physics and Astronomy, University of Delaware, Newark, DE 19716, USA}
\affiliation{Dept. of Physics, Yale University, New Haven, CT 06520, USA}
\affiliation{Columbia Astrophysics and Nevis Laboratories, Columbia University, New York, NY 10027, USA}
\affiliation{Dept. of Physics, University of Oxford, Parks Road, Oxford OX1 3PU, United Kingdom}
\affiliation{Dipartimento di Fisica e Astronomia Galileo Galilei, Universit{\`a} Degli Studi di Padova, I-35122 Padova PD, Italy}
\affiliation{Dept. of Physics, Drexel University, 3141 Chestnut Street, Philadelphia, PA 19104, USA}
\affiliation{Physics Department, South Dakota School of Mines and Technology, Rapid City, SD 57701, USA}
\affiliation{Dept. of Physics, University of Wisconsin, River Falls, WI 54022, USA}
\affiliation{Dept. of Physics and Astronomy, University of Rochester, Rochester, NY 14627, USA}
\affiliation{Department of Physics and Astronomy, University of Utah, Salt Lake City, UT 84112, USA}
\affiliation{Dept. of Physics, Chung-Ang University, Seoul 06974, Republic of Korea}
\affiliation{Oskar Klein Centre and Dept. of Physics, Stockholm University, SE-10691 Stockholm, Sweden}
\affiliation{Dept. of Physics and Astronomy, Stony Brook University, Stony Brook, NY 11794-3800, USA}
\affiliation{Dept. of Physics, Sungkyunkwan University, Suwon 16419, Republic of Korea}
\affiliation{Institute of Physics, Academia Sinica, Taipei, 11529, Taiwan}
\affiliation{Dept. of Physics and Astronomy, University of Alabama, Tuscaloosa, AL 35487, USA}
\affiliation{Dept. of Astronomy and Astrophysics, Pennsylvania State University, University Park, PA 16802, USA}
\affiliation{Dept. of Physics, Pennsylvania State University, University Park, PA 16802, USA}
\affiliation{Dept. of Physics and Astronomy, Uppsala University, Box 516, SE-75120 Uppsala, Sweden}
\affiliation{Dept. of Physics, University of Wuppertal, D-42119 Wuppertal, Germany}
\affiliation{Deutsches Elektronen-Synchrotron DESY, Platanenallee 6, D-15738 Zeuthen, Germany}

\author{R. Abbasi\,\orcidlink{0000-0001-6141-4205}}
\affiliation{Department of Physics, Loyola University Chicago, Chicago, IL 60660, USA}
\author{M. Ackermann\,\orcidlink{0000-0001-8952-588X}}
\affiliation{Deutsches Elektronen-Synchrotron DESY, Platanenallee 6, D-15738 Zeuthen, Germany}
\author{J. Adams}
\affiliation{Dept. of Physics and Astronomy, University of Canterbury, Private Bag 4800, Christchurch, New Zealand}
\author{J. A. Aguilar\,\orcidlink{0000-0003-2252-9514}}
\affiliation{Universit{\'e} Libre de Bruxelles, Science Faculty CP230, B-1050 Brussels, Belgium}
\author{M. Ahlers\,\orcidlink{0000-0003-0709-5631}}
\affiliation{Niels Bohr Institute, University of Copenhagen, DK-2100 Copenhagen, Denmark}
\author{S. Ali\,\orcidlink{0009-0001-2444-4162}}
\affiliation{Dept. of Physics and Astronomy, University of Kansas, Lawrence, KS 66045, USA}
\author{N. M. Amin}
\affiliation{Bartol Research Institute and Dept. of Physics and Astronomy, University of Delaware, Newark, DE 19716, USA}
\author{K. Andeen\,\orcidlink{0000-0001-9394-0007}}
\affiliation{Department of Physics, Marquette University, Milwaukee, WI 53201, USA}
\author{C. Arg{\"u}elles\,\orcidlink{0000-0003-4186-4182}}
\affiliation{Department of Physics and Laboratory for Particle Physics and Cosmology, Harvard University, Cambridge, MA 02138, USA}
\author{S. Athanasiadou}
\affiliation{Deutsches Elektronen-Synchrotron DESY, Platanenallee 6, D-15738 Zeuthen, Germany}
% ...

%% file: acknowledgments.tex
\begin{acknowledgements}
The authors gratefully acknowledge the support from the following agencies and institutions:
USA {\textendash} U.S. National Science Foundation-Office of Polar Programs,
U.S. National Science Foundation-Physics Division,
U.S. National Science Foundation-EPSCoR,
U.S. National Science Foundation-Office of Advanced Cyberinfrastructure,
Wisconsin Alumni Research Foundation,
Center for High Throughput Computing (CHTC) at the University of Wisconsin{\textendash}Madison,
Open Science Grid (OSG),
Partnership to Advance Throughput Computing (PATh),
Advanced Cyberinfrastructure Coordination Ecosystem: Services {\&} Support (ACCESS),
Frontera and Ranch computing project at the Texas Advanced Computing Center,
U.S. Department of Energy-National Energy Research Scientific Computing Center,
Particle astrophysics research computing center at the University of Maryland,
Michigan State University,
Astroparticle physics computational facility at Marquette University,
NVIDIA Corporation,
and Google Cloud Platform;
Belgium {\textendash} Funds for Scientific Research (FRS-FNRS and FWO),
FWO Odysseus and Big Science programmes,
and Belgian Federal Science Policy Office (Belspo);
Germany {\textendash} Bundesministerium f{\"u}r Forschung, Technologie und Raumfahrt (BMFTR),
Deutsche Forschungsgemeinschaft (DFG),
Helmholtz Alliance for Astroparticle Physics (HAP),
Initiative and Networking Fund of the Helmholtz Association,
Deutsches Elektronen Synchrotron (DESY),
and High Performance Computing cluster of the RWTH Aachen;
Sweden {\textendash} Swedish Research Council,
Swedish Polar Research Secretariat,
National Academic Infrastructure for Supercomputing in Sweden (NAISS),
and Knut and Alice Wallenberg Foundation;
European Union {\textendash} EGI Advanced Computing for research;
Australia {\textendash} Australian Research Council;
Canada {\textendash} Natural Sciences and Engineering Research Council of Canada,
Calcul Qu{\'e}bec, Compute Ontario, Canada Foundation for Innovation, WestGrid, and Digital Research Alliance of Canada;
Denmark {\textendash} Villum Fonden, Carlsberg Foundation, and European Commission;
New Zealand {\textendash} Marsden Fund;
Japan {\textendash} Japan Society for Promotion of Science (JSPS), Ministry of Education, Culture, Sports, Science and Technology (MEXT), and Institute for Global Prominent Research (IGPR) of Chiba University;
Korea {\textendash} National Research Foundation of Korea (NRF);
Switzerland {\textendash} Swiss National Science Foundation (SNSF).
\end{acknowledgements}